\documentclass[aps,prc,preprint,showpacs,floatfix]{revtex4}
\usepackage{amssymb}
\usepackage{graphicx}
\usepackage[colorlinks=true,linkcolor=blue]{hyperref}
\usepackage{latexsym}
\makeatother \makeatletter
\usepackage[T1]{fontenc}
\makeatletter
\usepackage{latexsym}
\usepackage{amsmath}

\begin{document}


\title{Dynamical instabilities for model with point-coupling interactions: Vlasov formalism method }

\author{Yanjun Chen$^{1,2}$}
\email{chenyjy@ustc.edu.cn}

\author{Juan Zou$^1$}
\author{Zipeng Cheng$^1$}
\author{Binguang He$^1$}
\affiliation{$^1$ Department of Physics and Electronic Science,
Changsha University of Science and Technology, Changsha, Hunan
410114, China\\ $^2$ Hunan Provincial Key Laboratory of Flexible
Electronic Materials Genome Engineering, Changsha, Hunan 410114,
China }
\renewcommand{\thefootnote}{\fnsymbol{footnote}}

\begin{abstract}

We explore the effects of the density dependence of symmetry energy
on the dynamical instabilities and crust-core phase transition in
the cold and warm neutron stars in the RMF theory with
point-coupling interactions using the Vlasov approach. The role of
the temperature and neutrino trapping has also been considered. The
distillation effect, crust-core transition density and pressure, the
cluster size and growth rates have been discussed. The present work
shows that the slope of symmetry energy, temperature, and neutrino
trapping have obvious effects.


\noindent

\end{abstract}

\maketitle

\section{ Introduction}

Neutron stars (NS) consist of a solid crust at low densities and a
homogeneous core in liquid phase. It is known that the uniform
liquid becomes unstable against small-amplitude density oscillations
when the density decreases from the high-density homogeneous core to
the inhomogeneous crust. Consequently, the phase transitions occur
which are associated to the liquid-gas phase transition in
asymmetric nuclear matter in the presence of electrons. It is
indicated that the properties of the crust and core-crust phase
transition play the important role in understanding some
astrophysical observations~\cite{link99,and12,cha13,lat13,pie14}.

It is well known that the relativistic mean field (RMF) theory can
successfully describe many nuclear phenomena and explain the
saturation mechanism of nuclear matter and the strong spin-orbit
interaction in finite nuclei in a consistent
way~\cite{serot86,rein89,ring96,bend03,meng06}. In recent years,
instead of the traditional RMF theory that is based on the effective
interaction between Dirac nucleons via the exchange of mesons, the
RMF model with point-coupling (PC)
interactions~\cite{bu02,ni08,zhao10}, which neglects mesonic degrees
of freedom and considers only interactions with zero range, has
become an alternative approach for the description of nuclear matter
and finite nuclei. It allows a simpler treatment of exchange terms
to study the effects beyond the mean-field for nuclear low-lying
collective excited states, and provides more opportunities to
investigate the relationship to the
non-relativistic approaches. 

Ref.~\cite{mou10} has used a widely used density-dependent
parametrization of PC model, DD-PC1~\cite{ni08}, to study the
core-crust transition density at zero temperature within the
thermodynamical approach and the effects of the density dependence
of symmetry energy. In this work, based on the parametrization
DD-PC1, we use the Vlasov formalism method, which is widely used to
study the stability of NS matter and the core-crust transition
within various
models~\cite{bri060,bri061,san08,du08,pais09,pais10,pais161,pais162},
to investigate the effects of the density dependence of the symmetry
energy on the dynamical instabilities and phase transitions expected
in the neutron star crust, considering the influences of finite
temperature and
neutrino trapping as well. 
The spectrum of collective modes, or the dispersion relation of the
system in the Vlasov approach arises from small oscillations around
the equilibrium state. The unstable collective modes are
characterized by an imaginary frequency of the dispersion relation.
Thus, the Vlasov formalism method, which incorporates the surface
and Coulomb effects, is more realistic than the thermodynamical one
which corresponds to the situation with the wave vector of
collective modes tending to zero.

This article is organized as follows. In Sec. II, we describe the
formulas necessary for the present work. In Sec. III, the calculated
results and some discussions are given. Finally, the summary is
present in Sec. IV.

\section{ The Formalism}

In this work, the Lagrangian for DD-PC1 parametrization reads
\begin{eqnarray}
\label{nld0}L=&&\bar{\psi}[\gamma^{\mu}(i\partial_{\mu}-eA_{\mu}\frac{1-\tau_3}{2})-M]\psi-\frac{1}{2}\alpha_S(\hat{\rho})(\bar{\psi}\psi)(\bar{\psi}\psi)-\frac{1}{2}\delta_S(\partial_{\nu}\bar{\psi}\psi)(\partial^{\nu}\bar{\psi}\psi)\nonumber\\
&&{}-\frac{1}{2}\alpha_V(\hat{\rho})(\bar{\psi}\gamma^{\mu}\psi)(\bar{\psi}\gamma_{\mu}\psi)-\frac{1}{2}\alpha_{tV}(\hat{\rho})(\bar{\psi}\vec{\tau}\gamma^{\mu}\psi)\cdot(\bar{\psi}\vec{\tau}\gamma_{\mu}\psi)\,\,,
\end{eqnarray}
where $\psi$ is the Dirac spinor of baryons and $A_{\mu}$ denotes
the electromagnetic field. The coupling parameter $\delta$ is
considered to be constant, while $\alpha$'s in the various
spin-isospin channels are analytical functions with respect to the
baryonic density $\rho$ alone, given by
$\alpha_i=a_i+(b_i+c_ix)e^{-d_ix}$ in which $x=\rho/\rho_{sat}$, and
$\rho_{sat}$ denotes the saturation density of symmetric nuclear
matter. The original DD-PC1 parameter sets are listed in Table
\ref{table0}.

The effective one-body Hamiltonian can be given by
\begin{eqnarray}
\label{nld1}
h_i=\left\{\begin{array}{cc}\sqrt{M^{*2}+(\vec{p}-\vec{V}_i)^2}+V^0_{i}\,\,,&\textrm{for}\;i=p,n\,\,,\\
\sqrt{m_e^{2}+(\vec{p}+e\vec{A})^2}-eA_0\,\,,&\textrm{for}\;i=e\,\,,\end{array}\right.
\end{eqnarray}
where the nucleon effective mass is defined as $M^*=M+\Sigma_S$, and
the scalar and vector self-energies, $\Sigma_S$ and
$V^{\mu}=\Sigma^{\mu}_V+\Sigma^{\mu}_R$, in which the rearrangement
term $\Sigma^{\mu}_R$ arises from the variation of the
density-dependent vertex functionals with respect to the nucleon
fields in the density operators, can be given by 
\begin{eqnarray}
\label{nld2}
\Sigma_S&=&\alpha_S\rho_S-\delta_S\triangle\rho_S\,\,,\\
\Sigma^{\mu}_{V}&=&\alpha_Vj^{\mu}+\alpha_{tV}\tau_ij^{\mu}_3+e\frac{1+\tau_i}{2}A^{\mu}\,\,,\\
\Sigma_R^{\mu}&=&(\alpha'_S\rho_S^2+\alpha'_Vj_{\nu}j^{\nu}+\alpha'_{tV}j_{3\nu}j_3^{\nu})j^{\mu}/2\rho_V\,\,,
\end{eqnarray}

The Vlasov equation describes the time evolution of the one-body
phase-space distribution functions for protons, neutrons, and
electrons, denoted by $f_{i\pm}(\vec{r},\vec{p},t)$, as
\begin{equation}
\label{nld3} \frac{\partial f_{i\pm}}{\partial
t}+\{f_{i\pm},h_{i\pm}\}=0\,\,,\,\,i=p,n,e,
\end{equation}
where +(--) denotes particles (antiparticles), and \{ , \} denotes
the Poisson brackets. 
Small deviations of the
distribution functions $\delta f$ around the equilibrium state can
be obtained with generating functions $S$ as
\begin{equation}
\label{nld4} \delta
f_{i\pm}=\{S_{i\pm},f_{0i\pm}\}=\{S_{i\pm},p^2\}\frac{df_{0i\pm}}{dp^2}\,\,,
\end{equation}
where $f_{0i}$ are equilibrium distribution functions. Of particular
interest are the longitudinal modes, with momentum $\vec{k}$ and
frequency $\omega$, described by the ansatz
\begin{eqnarray} 
\label{nld5} \left(\begin{array}{c}S_{i\pm}(\vec{r},\vec{p},t)\\
\delta\Sigma_S, \delta V^{\mu}\\ \delta\rho_S, \delta j^{\mu},\delta j_3^{\mu}\\
\delta A^{\mu}\end{array}\right)=\left(\begin{array}{c}S^i_{\omega\pm}(p,\cos\theta)\\
\delta\Sigma^S_{\omega}, \delta V^{\mu}_{\omega}\\ \delta\rho^S_{\omega}, \delta j^{\mu}_{\omega},\delta j^{\mu}_{3\omega}\\
\delta A^{\mu}_{\omega}\end{array}\right)e^{i(\omega
t-\vec{k}\cdot\vec{r})}\,\,,
\end{eqnarray} 
where $\theta$ is the angle between $\vec{p}$ and $\vec{k}$. In
terms of the generating functions, the linearized Vlasov equations
for $\delta f$ can be obtained. After transforming the unknowns
$S_{\omega}$ to the density oscillations, we obtain
\begin{eqnarray}
\label{nld6} \delta\rho_{\omega i}&=&\frac{2}{(2\pi)^3}\int
d^3p(\delta f_{i+}-\delta f_{i-})\nonumber\\
&=&\frac{-1}{2\pi^2T}[M^*\delta\Sigma_{\omega}^{Si}(I^{1i}_{\omega
+}+I^{1i}_{\omega -})+\delta V^{0i}_{\omega}(I^{2i}_{\omega
+}-I^{2i}_{\omega -})-\frac{\omega}{k}\delta
V^{i}_{\omega}(I^{2i}_{\omega +}-I^{2i}_{\omega -})]\,\,,\,\,i=p,n
\end{eqnarray}
\begin{eqnarray}
\label{nld7} \delta\rho^S_{\omega i}&=&\frac{2}{(2\pi)^3}[\int
d^3p(\delta f_{i+}+\delta f_{i-})\frac{M^*}{\epsilon}+\int
d^3p(f_{0i+}+f_{0i-})\delta(\frac{M^*}{\epsilon})]\nonumber\\
&=&\frac{-M^*}{2\pi^2T}[M^*\delta\Sigma_{\omega}^{Si}(I^{0i}_{\omega
+}-I^{0i}_{\omega -})+\delta V^{0i}_{\omega}(I^{1i}_{\omega
+}+I^{1i}_{\omega -})-\frac{\omega}{k}\delta
V^{i}_{\omega}(I^{1i}_{\omega +}+I^{1i}_{\omega
-})]\nonumber\\
&&{}+\frac{2}{(2\pi)^3}\int
d^3p(f_{0i+}+f_{0i-})\delta(\frac{M^*}{\epsilon})\,\,,\,\,i=p,n
\end{eqnarray}
and
\begin{eqnarray}
\label{nld8} \delta\rho_{\omega e}=\frac{e\delta
A^0_{\omega}}{2\pi^2T}(I^{2e}_{\omega +}-I^{2e}_{\omega
-})(1-\omega^2/k^2)\,\,,\,\,
\end{eqnarray}
in which
\begin{eqnarray}
\label{nld90} (\omega^2-k^2)\delta
A^0_{\omega}=-e(\delta\rho_{\omega p}-\delta\rho_{\omega e})\,\,,
\end{eqnarray}
and
\begin{eqnarray}
\label{nld9}
I^{ni}_{\omega\mp}=\int^{\infty}_{m_i}\epsilon^nI_{\omega\mp}(\epsilon)f_{0i\mp}(f_{0i\mp}-1)d\epsilon\,\,,
\end{eqnarray}
with
\begin{eqnarray}
\label{nld10}
I_{\omega\mp}(\epsilon)=\int^{p/\epsilon}_{-p/\epsilon}dx\frac{x}{\omega/k\pm
x}\,\,.
\end{eqnarray}
Here $m_i$ means $M^*$ for nucleon and $m_e$ for electron, $\delta\rho^S_{\omega p}$ and $\delta\rho^S_{\omega n}$ are 
the amplitudes of the oscillating scalar densities of protons and
neutrons, respectively, and $\delta\rho_{\omega p}$,
$\delta\rho_{\omega n}$, and $\delta\rho_{\omega e}$ are the
amplitudes of the oscillating proton, neutron, and electron
densities. 
For zero temperature, Eqs. \ref{nld6}-\ref{nld10} can still be
applied in the present work when one replaces $I^{ni}_{\omega\mp}$
given by
\begin{eqnarray}
\label{nld11}
I^{ni}_{\omega-}=0\,\,,\frac{I^{ni}_{\omega+}}{T}\Rightarrow-\epsilon_{iF}^nI_{\omega+}(\epsilon_{iF})\,\,,
\end{eqnarray}
in which $T$ is the temperature and $\epsilon_F$ is the Fermi energy
at zero temperature. After some derivation, including the equation
of motion of photons where the protons and electrons are sources of
Coulomb potential, Eqs. \ref{nld6}-\ref{nld8} can easily be put into
a matrix form as
\begin{eqnarray}
\label{nld12} M(\omega)\left(\begin{array}{c}\delta\rho^S_{\omega p}\\
\delta\rho^S_{\omega n}\\ \delta\rho_{\omega p}\\
\delta\rho_{\omega n}\\\delta\rho_{\omega
e}\end{array}\right)=0\,\,,
\end{eqnarray}
Then the dispersion relation of collective modes is obtained from
the determinant of $M(\omega)$. Eqs. \ref{nld6}-\ref{nld11} are the
universal equations for the investigation within the Vlasov
approach. It is worth pointing out that for the zero-range PC RMF
models, Eqs. \ref{nld6}-\ref{nld7} are the functions with respect of
the oscillating scalar and baryon densities, thus can directly be
put into the matrix form as Eq. \ref{nld12}, while for the
finite-range meson-exchange RMF models, Eqs. \ref{nld6}-\ref{nld7}
are the functions with respect of the oscillating meson fields, and
one has to use the equations of motion of mesons to replace the
oscillating meson fields in Eqs. \ref{nld6}-\ref{nld7} by the
oscillating densities.

\section{ Results and Discussion}


In this work, the parameters of isoscalar channels in the Lagrangian
Eq. \ref{nld0} remain unchanged in order that the properties of the
saturated symmetric nuclear matter, namely, the saturation density,
the binding energy, and the compression modulus, are kept fixed. We
vary the density dependence of symmetry energy by adjusting the
parameters of isovector channels, i.e., $b_{tV}$ and $d_{tV}$, in
Eq. \ref{nld0} and Table \ref{table0}, while keeping the symmetry
energy at saturation unchanged which is 33 MeV for DD-PC1.

\begin{figure}[!htb]
\includegraphics[width=15cm]{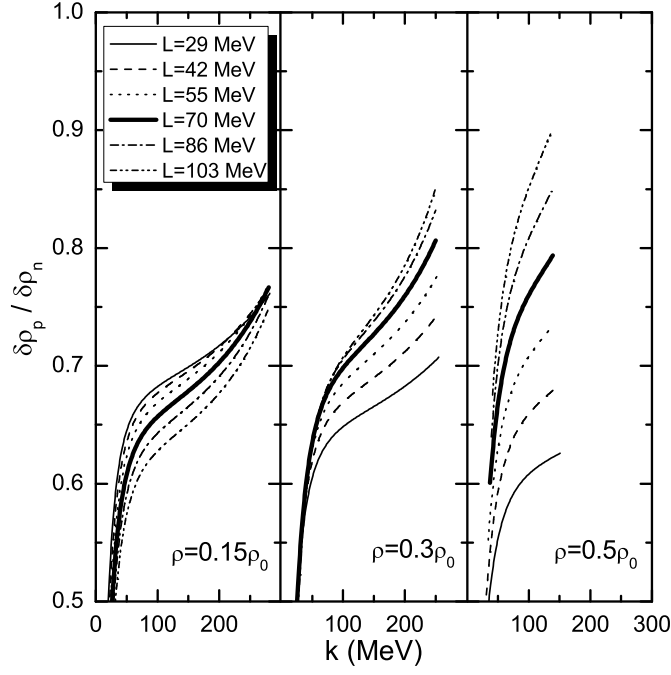}
\caption{\label{figg1}The ratio of the proton over neutron density
fluctuation $\delta\rho_p/\delta\rho_n$ at $T=4$ MeV plotted for the
proton fraction $y_p=0.3$, for $\rho=0.15\rho_0$, $0.3\rho_0$,
$0.5\rho_0$, as a function of the wave vector $\bf{k}$.
\textit{Thick curves} are for original DD-PC1 parametrization.}
\end{figure}

\begin{figure}[!htb]
\includegraphics[width=15cm]{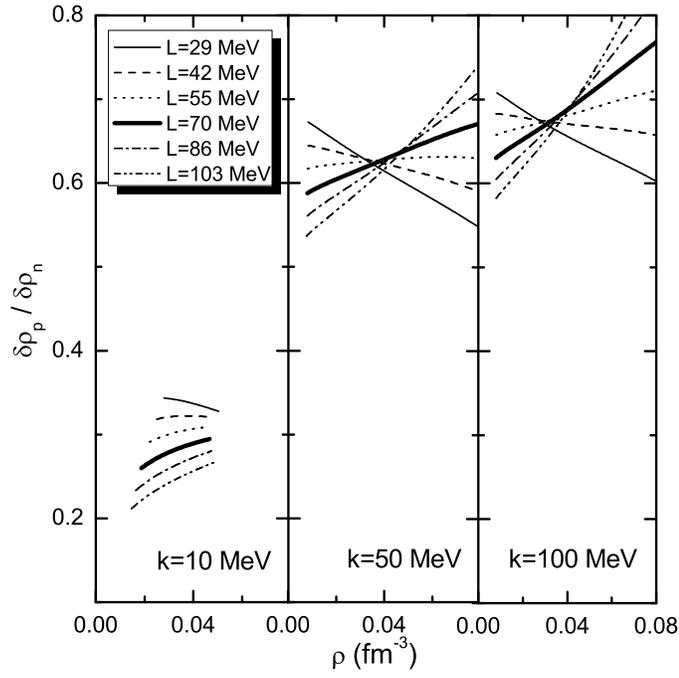}
\caption{\label{figg2}Same as Fig. \ref{figg1}, but for wave vector
k=10, 50, 100 MeV as a function of the density $\rho$. \textit{Thick
curves} are for original DD-PC1 parametrization.}
\end{figure}

It is known that the unstable modes correspond to the solutions of
the dispersion relation with imaginary frequencies $\omega=i\Gamma$,
where $\Gamma$ defines the exponential growth rate of the
instabilities. With these solutions, we study the instability
direction of the modes and the distillation effect, i.e., the denser
phase in nonhomogeneous nuclear matter prefers to be isospin
symmetric. We plot the ratio of the proton over neutron density
fluctuation $\delta\rho_p/\delta\rho_n$ as a function of the wave
vector $\vec{k}$ and the density $\rho$ in Fig. \ref{figg1} and Fig.
\ref{figg2}, respectively, for the proton fraction $y_p=0.3$ which
is close to that in $\beta$-equilibrium matter with neutrino
trapping at $T=4$ MeV. Fig. \ref{figg1} shows that for the largest
density considered, i.e., $\rho=0.5\rho_0$, the different slopes of
symmetry energy at saturation (parameter $L$) give rise to obviously
distinct results, specifically, large $L$ corresponding to large
distillation effect. With decreasing densities, the opposite
behavior is found. The phenomenon can be seen in Fig. \ref{figg2}
more obviously. Fig. \ref{figg2} shows that for $\rho\gtrsim 0.05$
fm$^{-3}$, the large $L$ increases the distillation effect, while at
lower densities, the opposite occurs, i.e., lower $L$ results in
larger $\delta\rho_p/\delta\rho_n$. It indicates that in the
nonhomogeneous region near the inner boundary of the crust, where
the densities are above a value, e.g., about 0.05 fm$^{-3}$ in this
case, the clusters prefer more proton rich for larger $L$, while in
the lower-density region of the crust, the larger $L$ prefers the
clusters with more neutron rich.


In Tables \ref{table1}-\ref{table2}, we show the transition
densities $n_t$ and corresponding pressures $P_t$ at the crust-core
transition in $\beta$-equilibrium neutron star matter for several
values of $L$, respectively.  The values in both tables are for
several temperatures with free
($Y_{\nu}=0$) and trapped ($Y_l=0.4$) neutrinos. 
The transition is defined as the crossing between the
$\beta$-equilibrium line and the spinodal surface. The
thermodynamical spinodal region requires the free energy curvature
matrix is negative, while the Vlasov spinodal surface corresponds to
the solutions of the dispersion relation with the frequency
$\omega=0$ and the moment $k=75$ MeV where the chosen value of $k$
in this work approximately defines the maximal spinodal region.
For the original DD-PC1 parametrization, the calculated $n_t$ for
$T=0$ MeV with thermodynamical method are 0.079 and 0.093 fm$^{-3}$
for neutrino-free and neutrino-trapped $\beta$-equilibrium matter,
respectively, while they  are about 10\% smaller for Vlasov
formalism method, which accordingly are 0.072 and 0.084 fm$^{-3}$,
respectively. The anticorrelation of $n_t$ and the slope $L$ has
been found in the literature using various
methods~\cite{oya07,xu09,du11,gri12,su14,bao14,bao15,wei18,gon19}.
Similarly, one sees in Table \ref{table1} that small $L$ corresponds
to the great value of
$n_t$. 
It can also be seen that $n_t$ decreases with temperature. Moreover,
in the crust of neutrino-free matter, we see that there is no
nonhomogeneous phase at temperatures above 4 MeV for $L>70$ MeV, and
even for very low $L$, no nonhomogeneous phase exists at $T=12$ MeV.
This mainly results from the fact that the spinodal region can
almost reach pure neutron matter at zero temperature while it is
more isospin symmetric for finite temperature (see Fig. 1 of Ref.
\cite{bri061}). Meanwhile, the proton fraction of
$\beta$-equilibrium neutrino-free matter is quite small at
subsaturation and the $\beta$-equilibrium line can only pass across
the spinodal region marginally. Thus the crust-core transition is
susceptible to the changes of temperature. It is known that the
greater $L$ corresponds to smaller symmetry energy at subsaturation
densities and favors more neutron-rich matter for homogeneous phase
at subsaturation. As a result, the nonhomogeneous phase can only
exist at low temperature for great $L$. As a contrast, the proton
fraction in the matter with trapped neutrinos is quite large,
$\sim$0.3. Therefore, Table \ref{table1} shows that the transition
densities do not differ much for various $L$ and the nonhomogeneous
phase still exists until $T=12$ MeV. It means that the
nonhomogeneous phase can exist at higher temperature in the crust of
the protoneutron star compared with that after neutrinos outflow. In
contrast with $n_t$, the dependence of $P_t$ on the slope $L$ is
nontrivial, as shown in Table \ref{table2}. At $T=0$ MeV, for
$Y_{\nu}=0$, $P_t$ increases with increasing $L$ for small $L$
region ($L\lesssim$ 55 MeV) and the opposite behavior occurs for
$L\gtrsim$ 55 MeV. This trend is similar to that observed in
Ref.~\cite{bao14,bao15,pais161}, while not to that in
Ref.~\cite{su14,gon19}. For $Y_l=0.4$, the trend for thermodynamical
method is different and $P_t$ increases monotonically with
increasing $L$. Moreover, it is observed that $P_t$ can become
downward with increasing $L$ when temperature increases. These
phenomenons might come from several completing effects, as discussed
in Ref.~\cite{bao14,bao15}, and can be model dependence.

\begin{figure}[!htb]
\includegraphics[width=15cm]{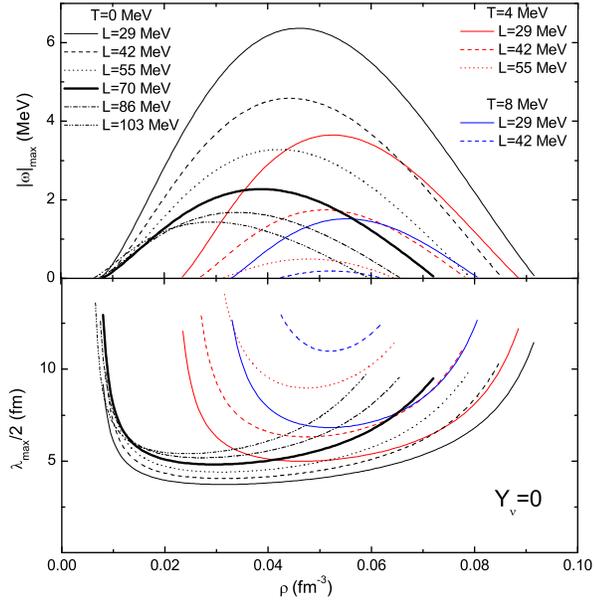}
\caption{\label{figg3}The growth rate of the most unstable modes
(\textit{upper panel}) and corresponding size of clusters
(\textit{lower panel}) as a function of density for
$\beta$-equilibrium neutrino-free matter $Y_{\nu}=0$. \textit{Thick
curves} are for original DD-PC1 parametrization.}
\end{figure}

\begin{figure}[!htb]
\includegraphics[width=15cm]{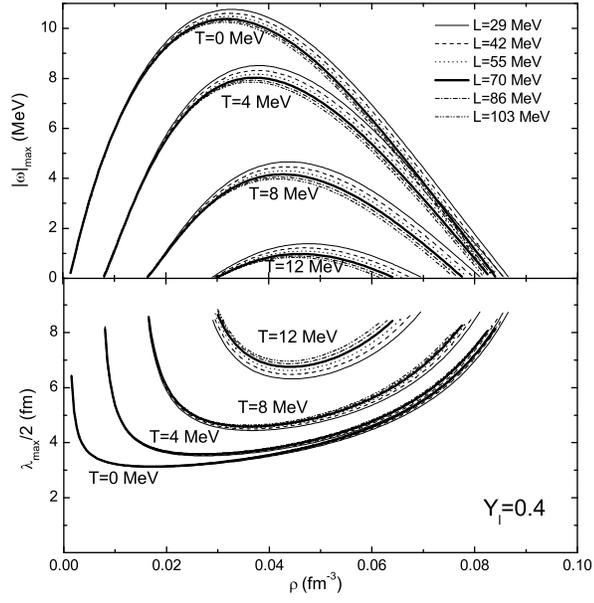}
\caption{\label{figg4}Same as Fig. \ref{figg3}, but for
$\beta$-equilibrium neutrino-trapped matter $Y_l=0.4$. \textit{Thick
curves} are for original DD-PC1 parametrization.}
\end{figure}

The most unstable mode is taken as the mode with the largest growth
rate $|\omega|_{max}$, which drives the matter to the nonhomogeneous
phase. Half of the wavelength $\lambda_{max}/2$ associated with this
mode is related to the most probable size of the clusters that are
formed by the perturbation. We plot $|\omega|_{max}$ (upper panels)
and corresponding $\lambda_{max}/2$ (lower panels) as a function of
density for $T=0$ MeV and several different finite temperatures,
$T=4$, 8, and 12 MeV in $\beta$-equilibrium matter without neutrinos
in Fig. \ref{figg3} and considering neutrino trapping with a lepton
fraction $Y_l=0.4$ in Fig. \ref{figg4}. We see from Fig. \ref{figg3}
that except for very low densities, e.g., $\rho\lesssim 0.02$
fm$^{-3}$, the smaller the slope $L$, the larger the growth rate and
the smaller the size of the clusters. It is seen that both the
largest value of the growth rates and the smallest size of the
clusters are shifted to larger densities when $L$ decreases.
Moreover, the density range for instabilities increases with
decreasing $L$. These phenomenons can also be seen in Fig.
\ref{figg4}. However, we see the differences between various $L$ in
Fig. \ref{figg4} are small. It can be explained by the large proton
fraction for matter with trapped neutrinos. The figures show that
the effects of the temperature are large and globally to reduce the
instability region and the growth rate and to increase the cluster
size. The largest growth rate and the smallest clusters are also
observed to shift to larger densities with increasing temperature.
For neutrino-free matter, there is no cluster in Fig. \ref{figg3} at
temperatures above 4 MeV for $L\gtrsim$ 70 MeV, while the clusters
still exist at $T=12$ MeV in Fig. \ref{figg4} for matter with
trapped neutrinos. Comparing these two figures, we see that the
neutrino trapping leads to larger growth rate and smaller clusters,
e.g., The smallest size of clusters is about 4 MeV for neutrino-free
matter, while $\thicksim$ 3 MeV when including neutrinos.


\section{ Summary}

In summary, based on the universal equations Eqs.
\ref{nld6}-\ref{nld11}, we have used the Vlasov formalism method to
explore the effects of the density dependence of symmetry energy on
the dynamical instabilities and crust-core phase transition in the
cold and warm neutron stars in the RMF theory with PC interactions.
The role of the temperature and neutrino trapping has also been
considered. We see that the clusters in the nonhomogeneous region
near the inner boundary of the crust prefer more proton rich for
larger $L$ at $\rho\gtrsim$ 0.05 fm$^{-3}$, while in the
lower-density region of the crust, the larger $L$ prefers the
clusters with more neutron rich. It is seen that $n_t$ decreases
when the slope $L$ or the temperature increases. For
$\beta$-equilibrium neutrino-free matter, the crust-core transition
is susceptible to the changes of temperature. The nonhomogeneous
phase can only exist at low temperature for great $L$. Even for very
low $L$, no nonhomogeneous phase exists at $T=12$ MeV. As a
contrast, the nonhomogeneous phase still exists until $T=12$ MeV in
the matter with trapped neutrinos and the transition densities do
not differ much for various $L$ due to the large proton fraction. In
contrast with $n_t$, the dependence of the transition pressure $P_t$
on the slope $L$ is nontrivial. At $T=0$ MeV, for $Y_{\nu}=0$, $P_t$
increases with increasing $L$ for $L\lesssim$ 55 MeV and the
opposite behavior occurs for $L\gtrsim$ 55 MeV, whereas for
$Y_l=0.4$, the trend for thermodynamical method is different which
increases monotonically with increasing $L$. When temperature
increases, $P_t$ can become downward with increasing $L$.

The slope of symmetry energy, temperature, and trapping neutrinos
have obvious effects on the estimated size of clusters and growth
rates. The small $L$ corresponds to small cluster size, large growth
rate and large density range for instabilities. The temperature
reduces the instability region and growth rate and increases the
cluster size. Moreover, the largest growth rate and the smallest
clusters are shifted to larger densities with decreasing $L$ or
increasing temperature. The neutrino trapping can reduce the effects
of $L$ due to the large proton fraction and leads to large growth
rate and small clusters.


\begin{acknowledgments}
The authors would like to thank the anonymous referee for her/his
constructive suggestions which are very helpful to improve this
manuscript. 
\end{acknowledgments}

\newpage
\begin{table}\caption{\label{table0}The parameters of the Lagrangian for the original DD-PC1 parametrization with parameter $\delta_S=-0.8149$ [fm$^{-4}$].}
\begin{tabular}{ccccc}\hline\hline &$a_i$ [fm$^{-2}$]&$b_i$ [fm$^{-2}$]&$c_i$ [fm$^{-2}$]&$d_i$\\
\hline $i=S$ &$-10.0462$  &$-9.1504$  &$-6.4273$  &1.3724\\
\hline $i=V$ &5.9195 &8.8637  & &0.6584\\
\hline $i=tV$& &1.8360 & &0.6403\\
\hline\hline\end{tabular}
\end{table}

\newpage
\begin{table}\caption{\label{table1}The crust-core transition densities [fm$^{-3}$] for $\beta$-equilibrium neutrino-free matter ($Y_{\nu}=0$) and neutrino-trapped matter ($Y_l=0.4$) for several slopes $L$ and temperatures $T$ including $T=0$ MeV. The results in italic type are for thermodynamical method and the other results are for Vlasov one. Those calculated with the original DD-PC1 parametrization are in bold type.}
\begin{tabular}{||c||c|c|c|c|c|c||}\hline\hline &$L$ (MeV)&{\it{T=0 MeV}}&T=0 MeV&4 MeV&8 MeV&12 MeV\\
\hline &29&{\it{0.096}}&0.091&0.087&0.079&\\
\cline{2-7} &42&{\it{0.091}}&0.085&0.077&0.058&\\
\cline{2-7} &55&{\it{0.086}}&0.079&0.063&&\\
\cline{2-7} \raisebox{2.3ex}[0pt]{$Y_{\nu}=0$}&{\bf{70}}&{\bf{0.079}}&{\bf{0.072}}&&&\\
\cline{2-7} &86&{\it{0.073}}&0.065&&&\\
\cline{2-7} &103&{\it{0.066}}&0.059&&&\\
\hline &29&{\it{0.095}}&0.087&0.086&0.081&0.07\\
\cline{2-7} &42&{\it{0.094}}&0.086&0.085&0.08&0.068\\
\cline{2-7} &55&{\it{0.094}}&0.085&0.084&0.079&0.066\\
\cline{2-7} \raisebox{2.3ex}[0pt]{$Y_l=0.4$}&{\bf{70}}&{\bf{0.093}}&{\bf{0.084}}&{\bf{0.083}}&{\bf{0.078}}&{\bf{0.065}}\\
\cline{2-7} &86&{\it{0.093}}&0.084&0.083&0.077&0.063\\
\cline{2-7} &103&{\it{0.093}}&0.083&0.082&0.076&0.062\\
\hline\hline\end{tabular}
\end{table}
\newpage
\begin{table}\caption{\label{table2}The crust-core transition pressures [MeV fm$^{-3}$] for several slopes $L$ and temperatures $T$ including $T=0$ MeV. The results in italic type are for thermodynamical method and the other results are for Vlasov one. Those calculated with the original DD-PC1 parametrization are in bold type.}
\begin{tabular}{||c||c|c|c|c|c|c||}\hline\hline &$L$ (MeV)&{\it{T=0 MeV}}&T=0 MeV&4 MeV&8 MeV&12 MeV\\
\hline &29&{\it{0.265}}&0.225&0.275&0.42&\\
\cline{2-7} &42&{\it{0.417}}&0.349&0.346&0.378&\\
\cline{2-7} &55&{\it{0.489}}&0.393&0.284&&\\
\cline{2-7} \raisebox{2.3ex}[0pt]{$Y_{\nu}=0$}&{\bf{70}}&{\bf{0.485}}&{\bf{0.365}}&&&\\
\cline{2-7} &86&{\it{0.404}}&0.282&&&\\
\cline{2-7} &103&{\it{0.283}}&0.182&&&\\
\hline &29&{\it{1.113}}&0.93&1.004&1.161&1.242\\
\cline{2-7} &42&{\it{1.131}}&0.93&1.006&1.152&1.211\\
\cline{2-7} &55&{\it{1.148}}&0.934&1.007&1.144&1.182\\
\cline{2-7} \raisebox{2.3ex}[0pt]{$Y_l=0.4$}&{\bf{70}}&{\bf{1.167}}&{\bf{0.936}}&{\bf{1.007}}&{\bf{1.134}}&{\bf{1.151}}\\
\cline{2-7} &86&{\it{1.182}}&0.936&1.004&1.122&1.123\\
\cline{2-7} &103&{\it{1.192}}&0.932&0.998&1.108&1.096\\
\hline\hline\end{tabular}
\end{table}
\newpage

\newpage

\newpage

\newpage
\end{document}